\documentclass[aps,pre,superscriptaddress,twocolumn,showpacs,longbibliography]{revtex4}

\usepackage{graphicx,epstopdf,url}
\usepackage[colorlinks=true,urlcolor=blue,citecolor=blue,linkcolor=blue,urlcolor=blue]{hyperref}

\usepackage[active]{srcltx}

\begin{document}
\title{Thermal rectifier based on asymmetric interaction of molecular chain with thermostats}

\author{Alexander V. Savin}
\affiliation{ N.N. Semenov Federal Research Center for Chemical Physics,
Russian Academy of Science (FRCCP RAS), Moscow, 119991, Russia}
\affiliation{Plekhanov Russian University of Economics, Moscow, 117997 Russia}

\begin{abstract}
The model of thermal rectifier based on the asymmetry of interaction of the molecular chain ends 
with thermostats is proposed in this work. The rectification mechanism is not related to the chain 
asymmetry, but to the asymmetry of the interaction of chain ends with thermostats, for instance, 
due to different lengths of the end thermostats. The chain can be homogeneous, it is only 
important that the thermal conductivity of the chain should depend on temperature strict 
monotonically. The effect is maximal when convergence of the thermal conductivity with 
increasing length just begins to manifest itself. 
In this case, the efficiency of thermal rectification can reach 25\%. 
These conditions are met for carbon nanoribbons and nanotubes. Therefore, they can be ideal
objects for the construction of thermal rectifiers based on the asymmetric interaction 
with thermostats. Numerical simulation of heat transfer shows that the rectification 
of heat transfer can reach 14\% for nanoribbons and 22\% for nanotubes.
\end{abstract}
\maketitle

\section{Introduction}

Thermal rectification (TR) is a phenomenon in which thermal transport along a specific axis is
dependent upon the sign of the temperature gradient or heat current (see review \cite{Roberts11}).
The thermal rectification has been studied earlier for anharmonic chains with two different
substrates \cite{Terraneo02,Li04,Hu06}, as well as for the Frenkel-Kontorova and Fermi-Pasta-Ulam
chains~\cite{Lan07}. In such cases, the introduced asymmetry of the two-segment structures
plays the role of the so-called thermal diode, breaking the left and right symmetry.
Chain anisotropy leading to TR can also be obtained by the non-uniform stretching of
the Lennard-Jones chain~\cite{Savin17}.

In this work, the possibility of another mechanism not related to the asymmetry of the chain
will be demonstrated. This mechanism ensures the rectification of heat transfer due to asymmetric
interaction with thermostats applied to chain edges, for instance, due to their different lengths
(see Fig.~\ref{fig01}). This mechanism works even if the chain itself is homogeneous. The only important
factor is a strict monotonous dependence of thermal conductivity of the chain on temperature.
The mechanism works only for those chain lengths and those temperatures for which a slow 
convergence of heat conductivity in the chain occurs, i.e. when thermal state of the chain is far 
from thermodynamic limit. Carbon nanotubes and nanoribbons can be used as such chains.
In 2006 Chang et al.~\cite{Chang06} have observed TR in the measurements of non-uniformly mass-loaded
carbon and boron nitride nanotubes. The TR mechanism proposed here makes it possible to explain
this experimental observation.

\section{1D Model}
To illustrate the proposed mechanism of TR let us consider 1D chain of rotators
with periodic potential of nearest-neighbor interaction~\cite{Giardina00,Gendelman00}.
Hamiltonian of this chain can be presented in the dimensionless form:
\begin{equation}
H=\sum_{n=1}^N \frac12\dot{\phi}_n^2+\sum_{n=1}^{N-1} V(\phi_{n+1}-\phi_n),
\label{f1}
\end{equation}
where $N$ is the number of molecules, $\phi_n$ is the rotation angle of the $n$-th molecule,
and $V(\phi)=1-\cos\phi$ is the potential of the nearest-neighbor interaction.
This chain has a finite thermal conductivity $\kappa$ for all temperatures $T>0$.
Thermal conductivity of the chain decreases monotonically with increasing temperature
($\kappa(T)\searrow 0$ for $T\nearrow\infty$).

Let us put $N_l$ left end chain particles in the Langevin thermostat with temperature $T_l$
and $N_r$ right end chain particles in the thermostat with temperature $T_r$.
In this case, the equations of motion for this system can be written in the form:
\begin{eqnarray}
\nonumber
\ddot{\phi}_1&=&V'(\phi_{2}-\phi_1)-\gamma_l\dot{\phi}_1+\xi_{1,l},\nonumber\\
\ddot{\phi}_n&=&V'(\phi_{n+1}-\phi_n)-V'(\phi_n-\phi_{n-1})-\gamma_l\dot{\phi}_n+\xi_{n,l},\nonumber\\
              &&~~1<n\le N_l,\nonumber\\
\ddot{\phi}_n&=&V'(\phi_{n+1}-\phi_n)-V'(\phi_n-\phi_{n-1}), \label{f2}\\
             &&~~N_l<n\le N-N_r,\nonumber\\
\ddot{\phi}_n&=&V'(\phi_{n+1}-\phi_n)-V'(\phi_n-\phi_{n-1})-\gamma_r\dot{\phi}_n+\xi_{n,r},\nonumber\\
             &&~~N-N_r< n< N,\nonumber\\
\ddot{\phi}_N&=&-V'(\phi_N-\phi_{N-1})-\gamma_r\dot{\phi}_N+\xi_{N,r}, \nonumber
\end{eqnarray}
where $\gamma_{l}=1/t_l$ and $\gamma_r=1/t_r$ are damping coefficients, $\xi_{n,l}$ and $\xi_{n,r}$
are normal random forces normalized according to the conditions
$\langle\xi_{n,\alpha}(t_1)\xi_{k,\alpha}(t_2)\rangle=2\gamma_\alpha T_\alpha\delta_{nk}\delta(t_2-t_1)$,
$\langle\xi_{n,l}(t_1)\xi_{k,r}(t_2)\rangle=0$,
$\alpha=l,r$.
\begin{figure}[tb]
\begin{center}
\includegraphics[angle=0, width=1.0\linewidth]{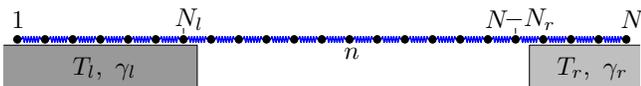}
\end{center}
\caption{\label{fig01}\protect
Schematic presentation of the chain of $N$ particles with ends asymmetrically interacting with 
Langevin thermostats at the temperatures $T_l$ and $T_r$ (the length of the left end $N_l$
is different from the length of the right end $N_r$).
Gray rectangles show Langevin thermostats with damping coefficients $\gamma_l$ and $\gamma_r$.
}
\end{figure}

Schematically, this chain is shown in Fig.~\ref{fig01}.
Damping coefficients characterize the intensity of the interaction of the edge chain particles
with thermostats. The substrate of the chain usually functions as a thermostat.
Therefore, the stronger is the interaction with the substrate, the greater is the value
of the coefficient $\gamma_\alpha$ (in general $\gamma_l\neq\gamma_r$).

\begin{figure}[tb]
\begin{center}
\includegraphics[angle=0, width=1.0\linewidth]{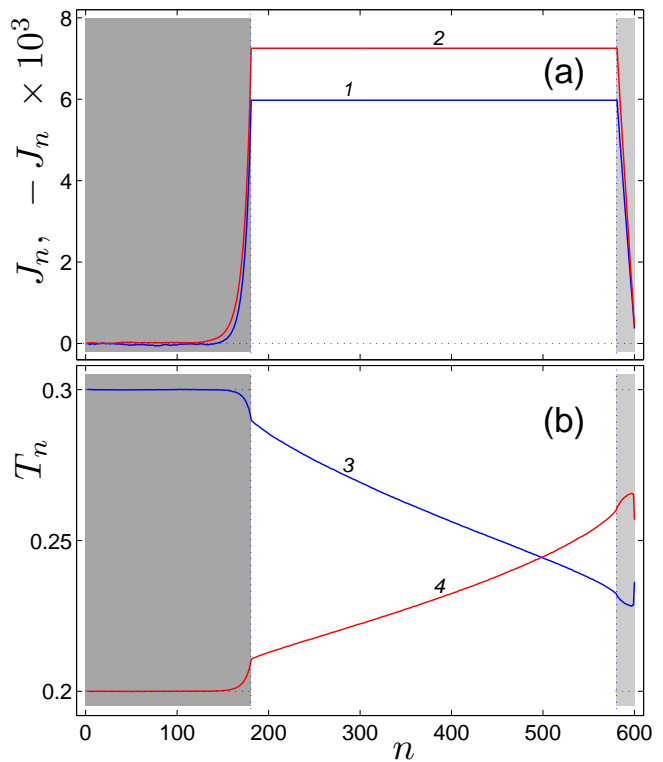}
\end{center}
\caption{\label{fig02}\protect
Distribution in the rotator chain of (a) local heat flow $J_n$ and (b) local temperatures $T_n$
for the thermostat temperatures $T_l=0.3$, $T_r=0.2$ and $T_l=0.2$, $T_r=0.3$ (curves 1, 3 and 2, 4).
Gray color marks the regions where the chain interacts with thermostats,
numbers $N_l=180$, $N_r=20$, damping coefficients $\gamma_l=0.1$, $\gamma_r=0.01$, length of the
chain $N=600$.
}
\end{figure}

Numerical simulation of the thermal transfer in this chain demonstrates that in the middle
region between the thermostats, $N_l<n\le N-N_r$, a local stationary flow of
heat is established, $J_n=-\langle\dot{\phi}_n V'(\phi_{n}-\phi_{n-1})\rangle\equiv J$,
that is characterized by a stationary profile of temperature $T_n=\langle\dot{\phi}^2_n\rangle$,
as shown in Fig.~\ref{fig02}.

To characterize the degree of anisotropy for describing the thermal flow, we take the edge temperatures
as $T_l=T_\pm$, $T_r=T_\mp$, where $T_\pm=T\pm0.05$ ($T$ is average temperature). Let $J_\pm$
be the heat flow caused by the temperature difference for $T_l=T_\pm$ and $T_r=T_\mp$ (heat
propagates from left to right when $J_+>0$, and in the opposite direction for $J_-<0$).
Then, anisotropy of the heat flow can be characterized by the heat anisotropy parameter defined as
$$
\varepsilon=\frac{J_++J_-}{J_+-J_-}.
$$
The anisotropy parameter takes the values $-1<\varepsilon< 1$; at $\varepsilon=0$ the anisotropy
vanishes ($J_+=-J_-$), and for $\varepsilon>0$ the thermal transfer is higher when heat propagates
from left to right than when it propagates from right to left ($J_+>-J_-$), and the reverse
otherwise when $\varepsilon<0$.
The anisotropy of heat transfer is often measured in percent:
$$
\eta=\left[\frac{\max(|J_+|,|J_-|)}{\min(|J_+|,|J_-|)}-1\right]\times100\%.
$$
In this case, the anisotropy $\eta$ can vary from zero to infinity.

The numerical simulation of heat transfer along the chain has demonstrated that the asymmetry
of the edge interactions with thermostats (asymmetry of thermostats) can result in up to 25\%
rectification of heat transfer. The mechanism of TR is clearly visible in Fig.~\ref{fig02}.
There is a greater temperature shift (downwards at $T_\alpha=T_+$, upwards at $T_\alpha=T_-$)
at the edge of the chain with weaker interaction in comparison to the other edge.
Thus, the average temperature of the chain during the heat transfer from a "stronger"\ thermostat
to a "weaker"\ one turns out to be higher than during heat transfer in the opposite direction.
Since the thermal conductivity of the rotators chain decreases sharply by increasing the
temperature, an upward shift of the temperature profile leads to a decrease of the heat flow,
while a downward shift leads to its increase. This results in the asymmetry of the heat transfer:
heat transfer from a "weaker"\ to a "stronger"\ thermostat is always higher than the heat transfer
in the opposite direction.
\begin{figure}[tb]
\begin{center}
\includegraphics[angle=0, width=1.0\linewidth]{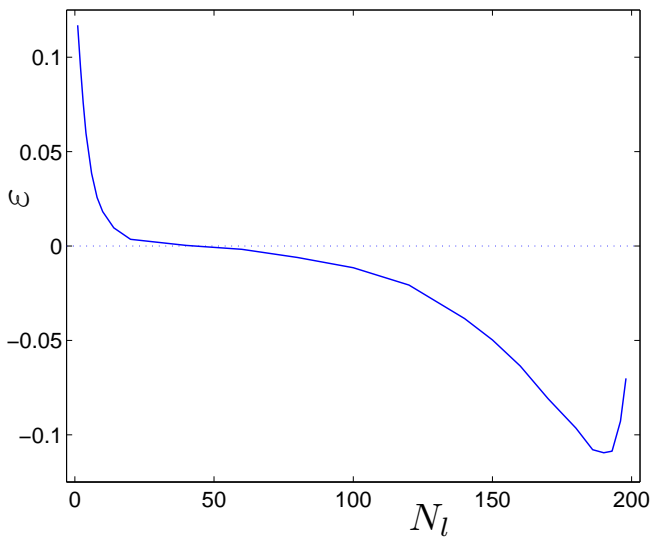}
\end{center}
\caption{\label{fig03}\protect
Dependence of heat transfer anisotropy $\varepsilon$ on the length of the left chain end interacting
with thermostat $N_l$ (the length of the right chain end $N_r= 200-N_l$) for the chain length $N=600$
(thermostat temperatures are $T_\pm=0.25\pm 0.05$, damping coefficients are $\gamma_l=0.1$, $\gamma_r=0.01$).
}
\end{figure}

The intensity of the interaction of chain end with a thermostat is determined by two parameters:
by the length of the end $N_\alpha$ and by the damping coefficient $\gamma_\alpha$ ($\alpha=l,r$).
The larger these parameters are, the stronger is the interaction of chain ends with thermostats.
Let us consider, for instance, the chain consisting of $N=600$ particles with the damping
coefficient $\gamma_l=0.1$ for the left edge and $\gamma_r=0.01$ for the right edge
(interaction of chain particles with the left substrate is ten times stronger than its interaction
with the right substrate). Let us assume that only 200 particles at the ends interact with thermostats
($N_l+N_r=200$), while the inner 400 particles do not interact with the substrates (thermostats).
Shifting the chain to the right or to the left (changing $N_l$), we can increase the interaction
with one thermostat and reduce it for the other one.

The dependence of the heat transfer anisotropy $\varepsilon$ on $N_l$ is shown in Fig.~\ref{fig03}.
As we can see from this figure, at the thermostat temperatures $T_\pm=0.25\pm 0.05$ the heat transfer
anisotropy is above zero when $N_l<40$ ($N_r>160$) and below zero when $N_l>40$ ($N_r<160$).
The decrease of $N_l$ (the increase of $N_r$) leads to a weakening of the left and strengthening
of the right thermostat (thermostats become "equal"\  when $N_l=40$, $N_r=160$),
while the increase of $N_l$ (the decrease of $N_r$) leads to the strengthening  of the left and
to the weakening of the right thermostat.
The anisotropy of heat transfer reaches the highest values at $N_l=1$, $N_r=199$
($\varepsilon=0.117$, $\eta=26\%$) and at $N_l=190$, $N_r=10$ ($\varepsilon=-0.110$, $\eta=25\%$).
\begin{figure}[tb]
\begin{center}
\includegraphics[angle=0, width=1.0\linewidth]{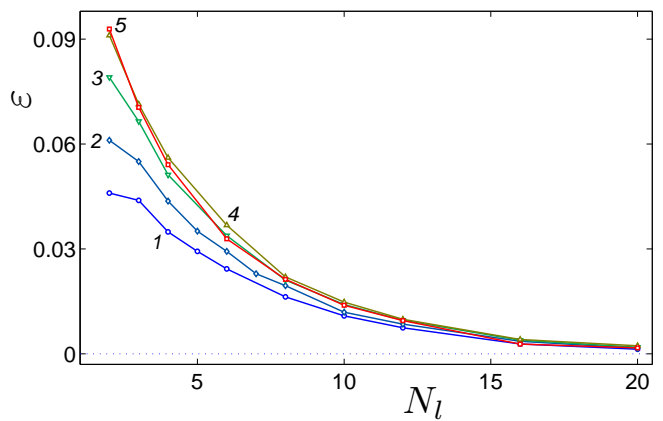}
\end{center}
\caption{\label{fig04}\protect
Dependence of heat transfer anisotropy $\varepsilon$ on the length of the left chain end $N_l$
interacting with the thermostat (length of the right end is $N_r=100-N_l$) for the chain consisting
of $N=150$, 200, 300, 500 and 900 particles (curves 1, 2, 3, 4 and 5). Thermostat temperatures are
$T_\pm=0.25\pm 0.05$, damping coefficients are $\gamma_l=\gamma_r=0.1$.
}
\end{figure}

If the damping coefficients are equal ($\gamma_l=\gamma_r=0.1$), the asymmetry of the interaction
with thermostats can be achieved by changing the lengths of the corresponding edge sections $N_l$
and $N_r$ of the chain. Let the total length of the edge sections $N_l+N_r= 100$ and temperatures
of the thermostats $T_\pm=0.25\pm0.05$. Let us examine how the change of the length of the left
chain edge $N_l$ affects on the anisotropy of heat transfer by different lengths of the chain $N$.
The dependence of the anisotropy $\varepsilon$ on $N_l$ at different lengths of the chain $N$
is presented in Fig.~\ref{fig04}. As we can see, the anisotropy of the heat transfer manifests
itself only when $N_l<20$. A further decrease of length leads to the monotonic increase of anisotropy.
For all chain lengths, the heat transfer anisotropy reaches its maximum at the minimum length
of the left edge. For $N\le 500$ the increase of chain length leads to the monotonic increase of anisotropy.
Anisotropy reaches its maximum value when the length $N=500$, while further increase of chain
length leads instead to the decrease of the heat transfer anisotropy.

\begin{figure}[tb]
\begin{center}
\includegraphics[angle=0, width=1.0\linewidth]{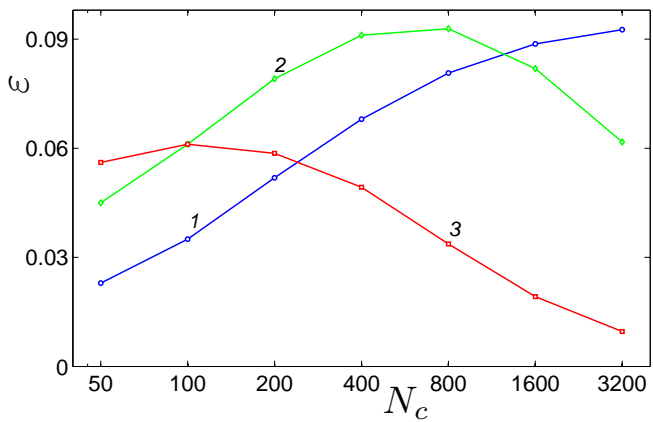}
\end{center}
\caption{\label{fig05}\protect
Dependence of heat transfer anisotropy $\varepsilon$ on the length of chain segment between
the end thermostats $N_c=N-N_l-N_r$ for temperature $T=0.15$, 0.25 and 0.35 ($T_\pm=T\pm0.05$)
-- curves 1, 2 and 3, correspondingly. The lengths of the end chain segments are $N_l=2$, $N_r=98$,
damping coefficients are $\gamma_l=\gamma_r=0.1$.
}
\end{figure}

The dependence of the maximum possible heat transfer anisotropy $\varepsilon$
on the chain length $N$ at different temperatures $T$ ($T_\pm=T\pm 0.05$) is shown in Fig.~\ref{fig05}.
As can be seen from the figure, for each temperature value there is its own optimal chain length
at which the anisotropy reaches its maximum value. For instance, at low temperature $T=0.15$
the maximum value $\varepsilon=0.093$ is reached when $N=3300$, at $T=0.25$ -- $\varepsilon=0.093$
when $N=900$, at $T=0.35$ -- $\varepsilon=0.061$ when N = 200.
The further increase of the chain length leads instead to a monotonic decrease of the heat
transfer anisotropy.

\begin{figure}[tb]
\begin{center}
\includegraphics[angle=0, width=1.0\linewidth]{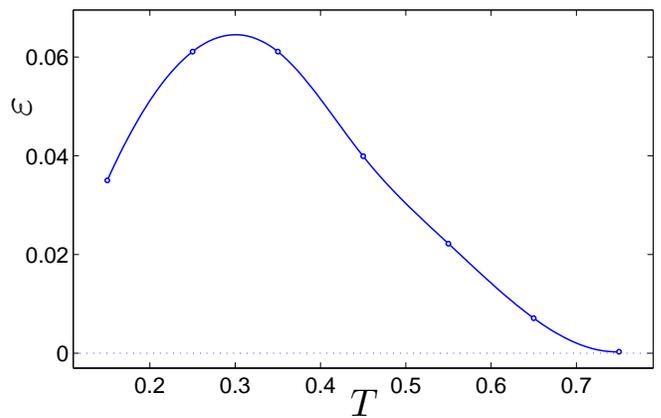}
\end{center}
\caption{\label{fig06}\protect
Dependence of heat transfer anisotropy $\varepsilon$ on temperature of the chain $T$
(temperatures of the end thermostats are $T_\pm=T\pm0.05$). Length of the chain is $N=200$,
lengths of the end segments are $N_l=2$, $N_r=98$, damping coefficients are $\gamma_l=\gamma_r=0.1$.
}
\end{figure}

The dependence of the maximum possible anisotropy of heat transfer on temperature $T$
for the chain of $N=200$ particles ($N_l=2$, $N_r=98$, $\gamma_l= \gamma_r=0.1$, $T_\pm=T\pm0.05$)
is presented in Fig.~\ref{fig06}. As can be seen from this figure, the increase of temperature
initially leads to increase of anisotropy. At $T=0.3$ the anisotropy reaches the maximum value
$\varepsilon=0.062$, and then decreases monotonically.
At temperatures $T>0.7$ the anisotropy of heat transfer becomes almost zero.

Thus, the asymmetry of edge thermostats at high temperatures does not lead to heat transfer
anisotropy due to the rapid convergence of heat conductivity.
The mechanism of heat transfer rectification based on the asymmetry of the edge thermostats
of the chain works only at low temperatures due to the slow convergence of heat conductivity.
The maximum 25\% value of heat transfer rectification can be reached only at chain lengths
for which the convergence of heat conduction only to begin manifesting itself,
i.e. at lengths comparable to the length of free path of long-wave phonons.

In carbon nanoribbons and nanotubes, long-wave phonons have a large free path length
and their thermal conductivity monotonically depends on temperature.
All this makes carbon nanoribbons and nanotubes ideal objects for the construction
of phonon rectifiers based on asymmetric interaction with thermostats.

\section{Carbon nanoribbons and nanotubes}

\begin{figure}[tb]
\begin{center}
\includegraphics[angle=0, width=1.0\linewidth]{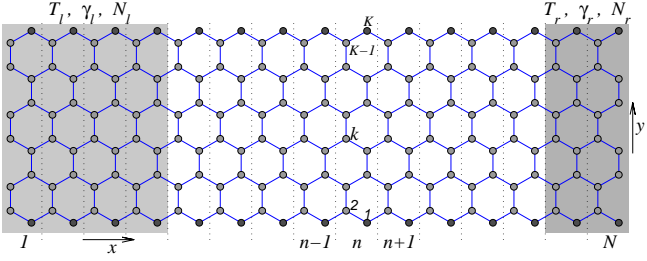}
\end{center}
\caption{\label{fig07}\protect
Full-atomic model of the carbon zigzag nanoribbon with the end asymmetrically interacting with the
Langevin thermostats at the temperatures $T_l$ and $T_r$. Nanoribbon lies on the $xy$ plane.
Gray regions mark the Langevin thermostats with damping coefficients
$\gamma_l$ and $\gamma_r$. The length of left thermostat $N_l$ is bigger than
that of the right one $N_r$, $N$ is the number of transverse elementary
cells of the nanoribbon, $K$ is the number of carbon atoms in each cell.
}
\end{figure}

\begin{figure}[tb]
\begin{center}
\includegraphics[angle=0, width=1.0\linewidth]{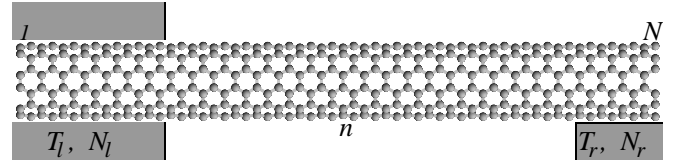}
\end{center}
\caption{\label{fig08}\protect
Full-atomic model of the carbon nanotube  with chirality index (6,6).
The cylindrical structure of the nanotube is formed by a longitudinal shift by a step
$a=2.46$~\AA\ of cyclic zigzag chains of the "armchair"\ structure consisting of $K=24$ carbon atoms.
The left edge of the nanotube containing  $N_l$ transverse cells is entirely embedded in the
bulk thermostat, the right edge containing $N_r$ cells lies on the flat substrate
(flat surface of the molecular crystal) that serves as the right thermostat
($T_l$ and $T_r$ -- temperature of left and right thermostat).
}
\end{figure}

Let us consider a finite flat carbon nanoribbon and nanotube with zigzag structure consisting of
$N\times K$ atoms -- see Fig.~\ref{fig07} and \ref{fig08} ($N$ is the number of transverse unit cells,
$K$ -- the number of atoms in the unit cell).
In the ground state the nanoribbon is flat. Initially, we assume that it lies in the
$xy$ plane and its symmetry center lies along the $x$ axis. Then its length can be calculated as
$L_x=(N-0.5)a$, width $L_y=3Kr_0/4-r_0$, where the longitudinal step of the nanoribbon is
$a=r_0\sqrt{3}$, $r_0=1.418$~\AA~ --  C--C valence bond length.

In realistic cases, the edges of the nanoribbon are always chemically modified. For simplicity,
we assume that the hydrogen atoms are attached to each edge carbon atom forming the edge line
of CH groups. In our numerical simulations, we take this into account by a change of the mass of
the edge atoms. We assume that the edge carbon atoms have the mass $M_1=13m_p$, while all other
internal carbon atoms have the mass $M_0=12m_p$, where $m_p=1.6601\times 10^{-27}$~kg is the proton
mass.

\begin{figure}[tb]
\begin{center}
\includegraphics[angle=0, width=1.0\linewidth]{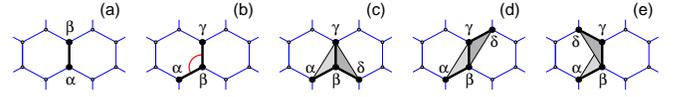}
\end{center}
\caption{\label{fig09}\protect
Configurations containing up to $i$th
type of nearest-neighbor interactions for (a) $i=1$, (b) $i=2$, (c) $i=3$,
(d) $i=4$, and (e) $i=5$.
}
\end{figure}

Hamiltonian of the nanoribbon and nanotube can be presented in the form,
\begin{equation}
H=\sum_{n=1}^{N}\sum_{k=1}^{K}\left[\frac12 M_{n,k}(\dot{\bf u}_{n,k},\dot{\bf u}_{n,k})+P_{n,k}\right],
\label{f3}
\end{equation}
where each carbon atom has a two-component index $\alpha=(n,k)$,
$n$ is the number of transversal elementary cell of zigzag nanoribbon (nanotube),
$k$ is the number of atoms in the cell. Here $M_\alpha$ is the mass
of the carbon atom with the index $\alpha$ (for internal atoms of nanoribbon and for all
atoms of nanotube, $M_\alpha=M_0$,
for the edge atoms of nanoribbon, $M_\alpha=M_1$), ${\bf u}_\alpha=(x_\alpha(t),y_\alpha(t),z_\alpha(t))$
is the three-dimensional vector describing the position of an atom with the index $\alpha$ at
the time moment $t$. The term $P_\alpha$ describes the interaction of the carbon atom with the
index $\alpha$ with the neighboring atoms.
The potential depends on variations in bond length, bond angles, and dihedral angles between
the planes formed by three neighboring carbon atoms and it can be written in the form
\begin{equation}
P=\sum_{\Omega_1}U_1+\sum_{\Omega_2}U_2+\sum_{\Omega_3}U_3+\sum_{\Omega_4}U_4+\sum_{\Omega_5}U_5,
\label{f4}
\end{equation}
where $\Omega_i$, with $i = 1$, 2, 3, 4, 5, are the sets of configurations including all
interactions of neighbors. This sets only need to contain configurations of the atoms shown
in Fig.~\ref{fig09}, including their rotated and mirrored versions.

Potential $U_1({\bf u}_\alpha,{\bf u}_\beta)$ describes the deformation energy due to
a direct interaction between pairs of atoms with the indexes $\alpha$ and $\beta$, as shown
in Fig.~\ref{fig09}(a). The potential $U_2({\bf u}_\alpha,{\bf u}_\beta,{\bf u}_\gamma)$
describes the deformation energy of the angle between the valence bonds ${\bf u}_\alpha,{\bf u}_\beta$
and ${\bf u}_\beta{\bf u}_\gamma$, see Fig.~\ref{fig09}(b).
Potentials $U_i({\bf u}_\alpha,{\bf u}_\beta,{\bf u}_\gamma,{\bf u}_\delta)$, $i = 3$, 4, and 5,
describes the deformation energy associated with a change in the angle between the planes
${\bf u}_\alpha,{\bf u}_\beta,{\bf u}_\gamma$ and ${\bf u}_\beta,{\bf u}_\gamma,{\bf u}_\delta$,
as shown in Figs.~\ref{fig09}(c)-(e).

We use the potentials employed in the modeling of the dynamics of large polymer
macromolecules \cite{Noid91,Sumpter94}:
for the valence bond coupling,
\begin{equation}
U_1({\bf u}_1,{\bf u}_2)\!=\!\epsilon_1
\{\exp[-\alpha_0(\rho-\rho_0)]-1\}^2,~\rho\!=\!|{\bf u}_2-{\bf u}_1|,
\label{f5}
\end{equation}
where $\epsilon_1=4.9632$~eV is the energy of the valence bond and $\rho_0=1.418$~\AA~
is the equilibrium length of the bond;
the potential of the valence angle
\begin{eqnarray}
U_2({\bf u}_1,{\bf u}_2,{\bf u}_3)=\epsilon_2(\cos\varphi-\cos\varphi_0)^2,~~
\label{f6}\\
\cos\varphi=({\bf u}_3-{\bf u}_2,{\bf u}_1-{\bf u}_2)/
(|{\bf u}_3-{\bf u}_2|\cdot |{\bf u}_2-{\bf u}_1|),~~
\nonumber
\end{eqnarray}
so that the equilibrium value of the angle is defined as $\cos\varphi_0=\cos(2\pi/3)=-1/2$;
the potential of the torsion angle
\begin{eqnarray}
\label{f7}
U_i({\bf u}_1,{\bf u}_2,{\bf u}_3,{\bf u}_4)=\epsilon_i(1+z_i\cos\phi),\\
\cos\phi=({\bf v}_1,{\bf v}_2)/(|{\bf v}_1|\cdot |{\bf v}_2|),\nonumber \\
{\bf v}_1=({\bf u}_2-{\bf u}_1)\times ({\bf u}_3-{\bf u}_2), \nonumber \\
{\bf v}_2=({\bf u}_3-{\bf u}_2)\times ({\bf u}_3-{\bf u}_4), \nonumber
\end{eqnarray}
where the sign $z_i=1$ for the indices $i=3,4$ (equilibrium value of the torsional angle $\phi_0=\pi$)
and $z_i=-1$ for the index $i=5$ ($\phi_0=0$).

The specific values of the parameters are $\alpha_0=1.7889$~\AA$^{-1}$,
$\epsilon_2=1.3143$ eV, and $\epsilon_3=0.499$ eV, they are found from the frequency
spectrum of small-amplitude oscillations of a sheet of graphite~\cite{Savin08}.
According to previous study~\cite{Gunlycke08}, the energy $\epsilon_4$ is close to
the energy $\epsilon_3$, whereas  $\epsilon_5\ll \epsilon_4$
($|\epsilon_5/\epsilon_4|<1/20$). Therefore, in what follows we use the values
$\epsilon_4=\epsilon_3=0.499$ eV and assume $\epsilon_5=0$, the latter means that we
omit the last term in the sum (\ref{f4}).

More detailed discussion and motivation of our choice of the interaction potentials
(\ref{f5}), (\ref{f6}), (\ref{f7}) can be found in earlier publication~\cite{Savin10}.

Let us consider $3K$-dimensional vector ${\bf x}_n=\{{\bf u}_{n,k}\}_{k=1}^K$
describing the positions of the atoms of the $n$-th cell. Then, the nanoribbon (nanotube)
Hamiltonian (\ref{f3}) can be written in the following form:
\begin{equation}
H\!=\!\!\sum_{n=2}^{N-1}\!h_n\!=\!\!\sum_{n=2}^{N-1}\!\!\left[
\frac12({\bf M}\dot{\bf x}_n,\dot{\bf x}_n)
\!+\!\!P({\bf x}_{n-1},{\bf x}_{n},{\bf x}_{n+1})\!\right]\!\!,
\label{f8}
\end{equation}
where the first term describes the kinetic energy of the atoms (${\bf M}$ is diagonal mass matrix
of the $n$-th elementary cell), and the second term describes the interaction between the atoms
in the cell and with the atoms of neighboring cells.

Hamiltonian (\ref{f8}) generates the system of equations of motion,
\begin{equation}
-{\bf M}\ddot{\bf x}_n=\frac{\partial~~}{\partial{\bf x}_n}H=F_n=P_{1,n+1}+P_{2,n}+P_{3,n-1},
\label{f9}
\end{equation}
where the function $P_{i,n}=P_i({\bf x}_{n-1},{\bf x}_{n},{\bf x}_{n+1})$,
$P_i=\partial P({\bf x}_{1},{\bf x}_{2},{\bf x}_{3})/\partial{\bf x}_i$,~$i=1,2,3$.

Local heat flux through the $n$-th cross section, $j_n$, determines a local longitudinal energy
density $h_n$ by means of a discrete continuity equation,
\begin{equation}
\dot{h}_n=j_n-j_{n-1}. \label{f10}
\end{equation}
Using the energy density from Eq. (\ref{f8}) and the motion equations (\ref{f9}),
we can derive the following relations:
\begin{eqnarray}
\dot{h}_n=({\bf M}\dot{\bf x}_n,\ddot{\bf x}_n)\!+\!(P_{1,n},\dot{\bf x}_{n-1})\!
+\!(P_{2,n},\dot{\bf x}_{n})\!+\!(P_{3,n},\dot{\bf x}_{n+1})\nonumber \\
=-(P_{1,n+1},\dot{\bf x}_{n})\!-\!(P_{3,n-1},\dot{\bf x}_{n})\!+\!(P_{1,n},\dot{\bf x}_{n-1})\!
+\!(P_{3,n},\dot{\bf x}_{n+1}).
\nonumber
\end{eqnarray}
From this and Eq. (\ref{f10}) it follows that the energy flux through $n$-th cross section of the
nanoribbon (nanotube) has the following simple form:
\begin{equation}
j_n=(P_{1,n},\dot{\bf x}_{n-1})-(P_{3,n-1},\dot{\bf x}_{n}).
\label{f11}
\end{equation}

\section{Interaction with thermostat}

In order to simulate asymmetric heat transfer in carbon nanoribbons and nanotubes,
it is necessary to estimate the intensity of their interaction with substrates,
which  will play role of external edge thermostats.
\begin{figure}[tb]
\begin{center}
\includegraphics[angle=0, width=1.0\linewidth]{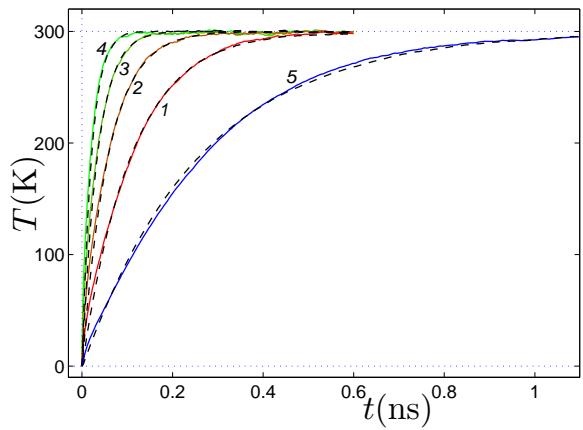}
\end{center}
\caption{\label{fig10}\protect
The time dependence of the temperature $T$ for nanoribbon lying on flat thermalized
graphite substrate for normalizing factor $c=1$, 2, 4, 8 (curves 1, 2, 3, 4).
Dashed lines show dependencies obtained through thermalization of isolated nanoribbon
using Langevin equations with relaxation time $t_0=109$, 58, 34, 21~ps.
Curve 5 shows the dependence for nanotube with chirality index (6,6)
(transversal cell has $K=24$ carbon atoms), dashed line shows the dependence for
isolated nanotube, in which only 10 atoms of each transversal cell interact
with Langevin thermostat with $t_0=109$~ps.
}
\end{figure}

The interaction of nanoribbons (nanotubes) with a thermostat is described
by the Langevin system of equations
\begin{equation}
{\bf M}\ddot{\bf x}_n=-F_n-\gamma{\bf M}\dot{\bf x}_n+\Xi_n,~~1\le n\le N,
\label{f12}
\end{equation}
where damping coefficient $\gamma=1/t_0$ ($t_0$ -- relaxation time) and $\Xi_n=\{\xi_{n,k,i}\}_{k=1,i=1}^{K,~3}$
is $3K$-dimensional vector of normally distributed random forces normalized by conditions
$$
\langle\xi_{n_1,k_1,i}(t_1)\xi_{n_2,k_2,j}(t_2)\rangle\!\!
=\!\! 2M_{n_1,k_1}\!\gamma k_BT\delta_{n_1n_2}\!\delta_{k_1k_2}\!\delta_{ij}\delta(t_1-t_2),
$$
where $k_B$ is the Boltzmann constant.
The intensity of the interaction with a thermostat is determined by the relaxation time $t_0$
of the velocity of the atom as the result of its interaction with the thermostat
(the shorter is the time $t_0$, the stronger is the interaction with the thermostat).

The role of thermostats is usually played by the substrate on which the nanoribbon (nanotube) lies.
Let us estimate the relaxation time $t_0$ for various substrates. For this purpose, let us consider
a two-layer nanoribbon of size $19.53\times 1.56$~nm$^2$ consisting of $2\times 1280$ carbon atoms
($N= 80$, $K=16$). We will consider the interaction of nanoribbons between themselves as the sum
of the pair interactions of their atoms. Non-valent pair interactions of carbon atoms for
nanoribbons and nanotubes can be adequately described with the help
of the Lennard-Jones potential \cite{Setton96}.
\begin{equation}
V(r)=c\epsilon_0[(r_0/r)^{12}-(r/r_0)^6],
\label{f13}
\end{equation}
where the bond energy $\epsilon_0=0.00276$~eV, the equilibrium bond length $r_0=3.809$\AA\
($c\ge 1$ is a normalizing factor allowing  to consider a stronger interaction).
\begin{figure}[tb]
\begin{center}
\includegraphics[angle=0, width=1.0\linewidth]{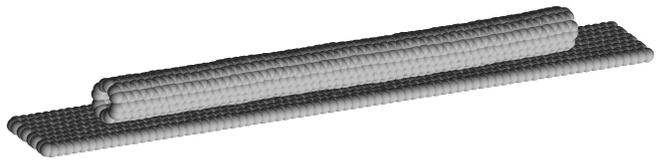}
\end{center}
\caption{\label{fig11}\protect
Carbon nanotube with chirality index (6,6) (number of transversal elementary
cells $N_1=60$, number of atoms in each cell $K_1=24$) lying on a flat graphene
nanoribbon consisting of $N_2\times K_2=80\times 16$ carbon atoms.
}
\end{figure}

Let us take a two-layer nanoribbon in the equilibrium position and place its first layer
in a Langevin thermostat with damping coefficient $\gamma=1/t_0$, $t_0= 0.05$~ps.
Let us then consider the thermalization of the second layer, which occurs through non-valent interactions.
To do so, we analyze the time dependence of the temperature for the second layer $T(t)$.
As can be seen in Fig.~\ref{fig10},  when $c=1$ the thermalization of the second layer occurs
as if we have thermalized only a single-layer nanoribbon using a Langevin thermostat with
relaxation time $t_0=109$~ps. When the interaction between the layers is increased by $c$ times
(in the interaction potential (\ref{f13}) factor $c>1$), the relaxation time decreases: for $c=2$,
time $t_0= 58$; for $c=4$, $t_0=34$; for $c=8$, $t_0=21$ and for $c=16$, time $t_0=13$~ps.
\begin{figure}[tb]
\begin{center}
\includegraphics[angle=0, width=1.0\linewidth]{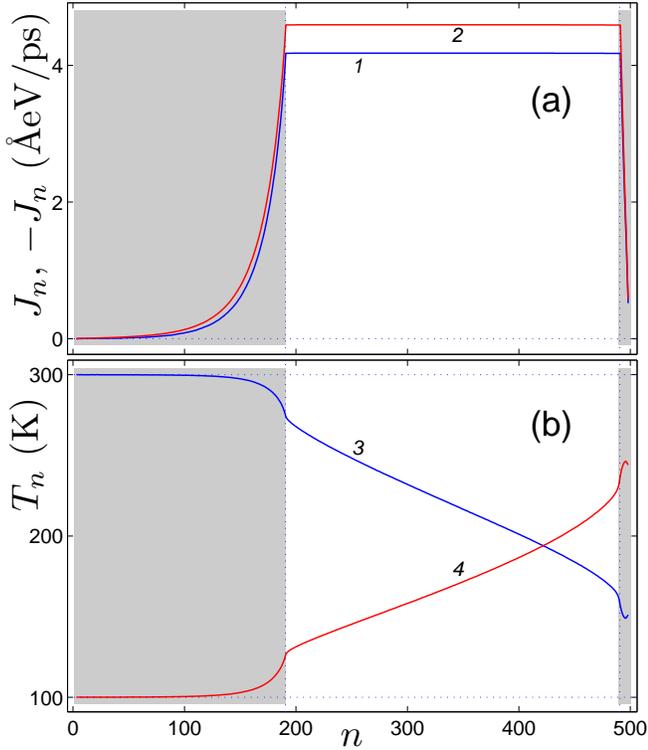}
\end{center}
\caption{\label{fig12}\protect
Distribution of (a) local heat flux $J_n$ and (b) local average temperature $T_n$ along the nanoribbon
of size $122.68\times 1.56$~nm$^2$ ($N=500$, $K=16$)
for the thermostat temperatures $T_l=300$K, $T_r=100$K
and $T_l=100$K, $T_r=300$K (curves 1, 3 and 2, 4).
Gray color marks the regions where the nanoribbon interacts with thermostats,
numbers $N_l=190$, $N_r=10$, damping coefficients $\gamma_l=\gamma_r=1$~ps$^{-1}$,
heat transfer anisotropy $\varepsilon=-0.047$.
}
\end{figure}
\begin{figure}[tb]
\begin{center}
\includegraphics[angle=0, width=1.0\linewidth]{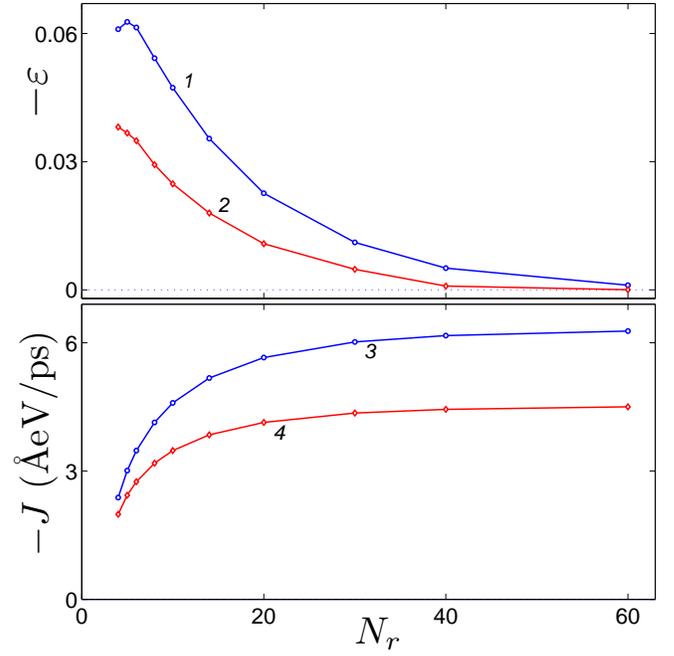}
\end{center}
\caption{\label{fig13}\protect
Dependence (a) of heat transfer anisotropy $\varepsilon$ and (b) of heat flow $J_-$
on the length of the right edge $N_r$ interacting with the thermostat
(length of the left edge $N_l=200-N_r$) for the nanoribbon of size
$122.68\times 1.56$~nm$^2$ ($N=500$, $K=16$) at thermostat temperatures
$T_\pm=200\pm 100$K (curves 1, 3) and $T_\pm=400\pm 100$K (curves 2, 4).
}
\end{figure}

The interaction of carbon atoms with nickel atoms can be described by the Morse potential
\begin{equation}
V(r)=\epsilon_0[e^{-\beta(r-r_0)} - 1]^2-\epsilon_0,
\label{f14}
\end{equation}
where the bond energy $\epsilon_0=0.433$~eV, the equilibrium bond length $r_0=2.316$\AA,
the parameter $\beta=3.244$\AA$^{- 1}$ \cite{Katin18}.
By calculating the energy of non-valent interaction of a carbon atom with the flat surface
of a graphite crystal we get the value $E_{\rm C}=0.052$~eV,
and for the flat surface of a nickel crystal -- the interaction energy $E_{\rm Ni}=0.8$~eV
(by the valence interaction the binding energy usually has several eV).
Therefore, the simulation of the thermalization of the second layer of nanoribbon allows us
to estimate the relaxation time for the Langevin thermostat: $t_0\approx 100$~ps
for weak non-valent interaction with the substrate formed by the surface of the molecular crystal
(graphite, silicon, silicon carbide), $t_0\approx 10$~ps for interaction with the flat surface
of the nickel crystal and $t_0\approx 1$~ps for the substrate with strong covalent interaction
with atoms of the nanoribbon.

When a nanotube lies on a flat substrate, only atoms adjacent to the substrate are interacting
with it. This is the reason why the thermalization of the nanotube should occur slower compared
to nanoribbons. To simulate this, let us consider the nanotube with chirality index (6,6)
lying on a flat substrate made of graphene nanoribbon with fixed edge atoms (see Fig.~\ref{fig11}).
We take the nanotube consisting of $N_1\times K_1=60\times 24$ atoms and a nanoribbon consisting
of $N_2\times K_2=80\times 16$ carbon atoms. Then we describe the interactions of nanotube atoms
with nanoribbon atoms using the Lennard-Jones pair potential (\ref{f13}) with the factor $c=1$.

Let us take the ground state of a two-component system of nanoribbon+nanotube and place
the nanoribbon into the Langevin thermostat with time relaxation $t_0=0.05$~ps.
Let us then consider the thermalization of the nanotube, which occurs through non-valence
interactions (factor $c=1$). In order to do so, we will analyze the time dependence of the nanotube
temperature $T(t)$. As can be seen in Fig.~\ref{fig10}, the thermalization of the nanotube occurs
as if we have put the isolated nanotube into the Langevin thermostat with relaxation time
$t_0=260$~ps (2.5 times slower than for a flat nanoribbon).
The same thermalization rate can be obtained if we take into account the interaction with
the Langevin thermostat with time relaxation $t_0=109$~ps for only 10 atoms in each transversal
cell of the nanotube. Thus, only 10 atoms of each transversal cell of the nanotube will
effectively interact with the flat substrate formed by the surface of a graphite crystal.

\section{Asymmetrical heat transfer along carbon nanoribbons}

Let us consider a carbon nanoribbon whose ends interact asymmetrically with Langevin thermostats
(see Fig.~\ref{fig07}). Let us take a nanoribbon in its ground state and fix the position of atoms
of its first ($n=1$) and last ($n=N$) transverse cells (fixed boundary conditions).
Then let us put its first $N_l$ transverse cells in the Langevin thermostat with temperature $T=T_l$
and damping coefficient $\gamma=\gamma_l$, while putting its last $N_r$ cells in the thermostat
with $T=T_r$, $\gamma=\gamma_r$.
In this case, the dynamics of the nanoribbon will be described by the Langevin system of equations
\begin{eqnarray}
{\bf M}\ddot{\bf x}_n&=&-F_n-\gamma_l{\bf M}\dot{\bf x}_n+\Xi_{n,l},~~1< n\le N_l,\nonumber \\
{\bf M}\ddot{\bf x}_n&=&-F_n,~~N_l<n\le N-N_r, \label{f15} \\
{\bf M}\ddot{\bf x}_n&=&-F_n-\gamma_r{\bf M}\dot{\bf x}_n+\Xi_{n,r},~~N-N_r< n< N, \nonumber
\end{eqnarray}
where $\Xi_{n,\alpha}=\{\xi_{n,k,i}\}_{k=1,i=1}^{K,~~3}$ is $3K$-dimensional vector of normally
distributed random forces normalized by conditions
\begin{eqnarray}
\langle\xi_{n_1,k_1,i,\alpha}(t_1)\xi_{n_2,k_2,j,\alpha}(t_2)\rangle=\nonumber\\
2\gamma_\alpha k_BT_\alpha M_{n_1,k_1}\delta_{n_1n_2}\delta_{k_1k_2}\delta_{ij}\delta(t_2-t_1),~\alpha=l,r,
\label{f16}\\
 \langle\xi_{n_1,k_1,i,l}(t_1)\xi_{n_2,k_2,j,r}(t_2)\rangle=0.\nonumber
\end{eqnarray}
\begin{figure}[tb]
\begin{center}
\includegraphics[angle=0, width=1.0\linewidth]{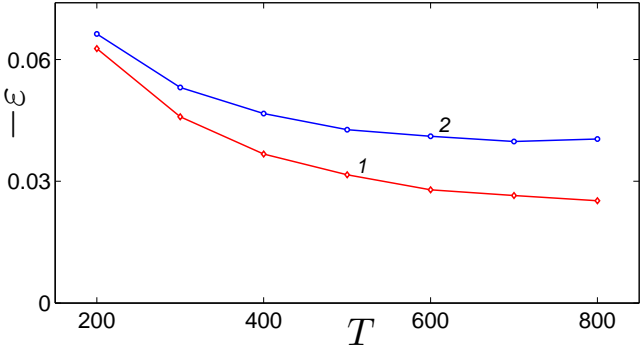}
\end{center}
\caption{\label{fig14}\protect
Dependence of the maximum possible heat transfer anisotropy $\varepsilon$ on
temperature $T$ for nanoribbon and nanotube (curves 1 and 2).
Number of transverse elementary cells $N=500$, thermostat temperatures $T_\pm=T\pm 100$K.
}
\end{figure}

For the convenience of numerical simulation, we take $\gamma_l=\gamma_r=1/t_0$ with relaxation
time $t_0=1$~ps (using large values of relaxation time requires a longer numerical simulation).
The asymmetry of the interaction of the nanoribbon edges with thermostats is ensured due
to the inequality of their lengths $N_l$ and $N_r$.

Let us consider a nanoribbon of size $122.68\times 1.56$~nm$^2$ ($N=500$, $K=16$)
with thermostat temperatures $T_l=T_\pm$, $T_r=T_\mp$, where $T_\pm=T\pm 100$K
($T$ is average temperature).
We select the initial conditions for system (\ref{f15}) corresponding to the ground state of the
nanoribbon, and solve the equations of motion numerically tracing the transition to the regime
with a stationary heat flux. At inner part of the nanoribbon $N_l<n<N-N_r$, we observe the formation
of a temperature gradient corresponding to a constant flux. Distribution of the average
values of temperature and heat flux along the nanoribbon can be found in the form
\begin{eqnarray*}
T_n=\lim_{t\rightarrow\infty}\frac{1}{3Kk_Bt}\int_0^t
({\bf M}\dot{\bf x}_n(\tau),\dot{\bf x}_n(\tau))d\tau,\\
J_n=\lim_{t\rightarrow\infty}\frac{a}{t}\int_0^tj_n(\tau)d\tau.
\end{eqnarray*}

Distribution of the temperature and local heat flux along the nanoribbon is shown in Fig.~\ref{fig12}.
The heat flux in each cross section of the inner part of the nanoribbon should remain constant,
namely, $J_n\equiv J$ for $N_l<n<N-N_r$. The requirement of independence of the heat flux $J_n$
on a local position $n$ is a good criterion for the accuracy of numerical simulations,
as well as it may be used to determine the integration time for calculating the mean values
of $J_n$ and $T_n$. As follows from the figure, the heat flux remains constant along the central
inner part of the nanoribbon.
\begin{figure}[tb]
\begin{center}
\includegraphics[angle=0, width=1.0\linewidth]{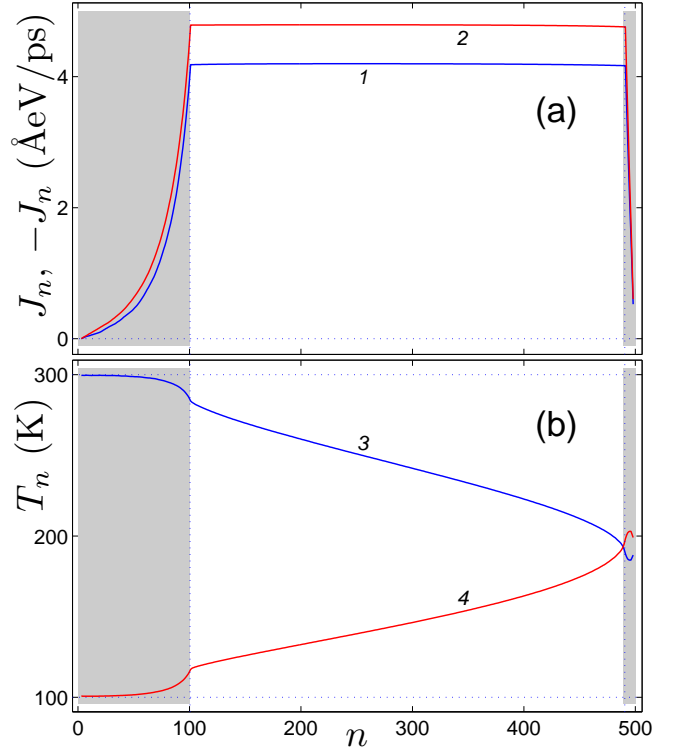}
\end{center}
\caption{\label{fig15}\protect
Distribution of (a) local heat flux $J_n$ and (b) local average temperature $T_n$ along the nanotube
of length  $L=122.68$~nm ($N=500$, $K=24$) for the thermostat temperatures $T_l=300$K, $T_r=100$K
and $T_l=100$K, $T_r=300$K (curves 1, 3 and 2, 4).
Gray color marks the regions where the nanotube interacts with thermostats,
numbers $N_l=100$, $N_r=10$, damping coefficients $\gamma_l=\gamma_r=1$~ps$^{-1}$,
heat transfer anisotropy $\varepsilon=-0.066$.
}
\end{figure}

Let us assume that only 200 edge transverse cells of the nanoribbon interact with thermostats
($N_l+N_r=200$), while the inner 300 cells never interact with edge thermostats (substrates).
By shifting the nanoribbon to the right or to the left (changing $N_r$) we can increase
or decrease the interaction with the thermostat of one edge and reduce or increase
the interaction for the other edge of the nanoribbon.
The dependence of the anisotropy of heat transfer $\varepsilon$ on $N_r$ is demonstrated in Fig.~\ref{fig13}.
As the figure shows, the anisotropy of heat transfer begins to manifest itself only when $N_r<60$.
The decrease of $N_r$ (the increase of $N_l$) leads to a monotonic increase of anisotropy.
Anisotropy reaches its maximum when $N_r=5$ (for thermostat temperatures $T_\pm=200\pm 100$~K
anisotropy $\varepsilon=-0.063 $, for $T_\pm=400\pm 100$ K -- $\varepsilon=-0.037$).
With the increase of the average temperature value $T$, the anisotropy of heat transfer
is monotonically weakening but remains significant for all values $T<900$~K -- see Fig.~\ref{fig14}.

\section{Asymmetrical heat transfer along carbon nanotubes}

Let us consider a carbon nanotube with chirality index (6,6) whose left edge (consisting of $N_l$
transverse elementary cells) is embedded in a volume thermostat, and whose right edge ($N_r$ cells)
lies on a flat substrate that functions as the right thermostat (see Fig.~\ref{fig08}).
Then the dynamics of the nanotube will be described by the Langevin system of equations
\begin{eqnarray}
M_0\ddot{\bf u}_{n,k}&=&-F_{n,k}-\gamma_l M_0\dot{\bf u}_{n,k}+\Xi_{n,k,l},
\nonumber\\
&~& 1<n\le N_l,~~1\le k\le K, \nonumber \\
M_0\ddot{\bf u}_{n,k}&=&-F_{n,k},\label{f17}\\
&~& N_l<n\le N-N_r,~~1\le k\le K,\nonumber\\
M_0\ddot{\bf u}_{n,k}&=&-F_{n,k}-\gamma_r M_0\dot{\bf u}_{n,k}+\Xi_{n,k,r},
\nonumber\\
&~& N-N_r<n< N,~1\le k\le 10,\nonumber\\
M_0\ddot{\bf u}_{n,k}&=&-F_{n,k},\nonumber\\
&~&N-N_r<n<N,~~11\le k\le K, \nonumber
\end{eqnarray}
where force $F_{n,k}=\partial H/\partial{\bf u}_{n,k}$, $K=24$,
damping coefficient $\gamma_l=\gamma_r=1/t_0$ (relaxation time $t_0=1$~ps)
and $\Xi_{n,k,\alpha}=\{\xi_{n,k,i,\alpha}\}_{i=1}^3$, index $\alpha=r,l$, is 3-dimensional
vector of normally distributed random forces  normalized by conditions (\ref{f16}).
\begin{figure}[tb]
\begin{center}
\includegraphics[angle=0, width=1.0\linewidth]{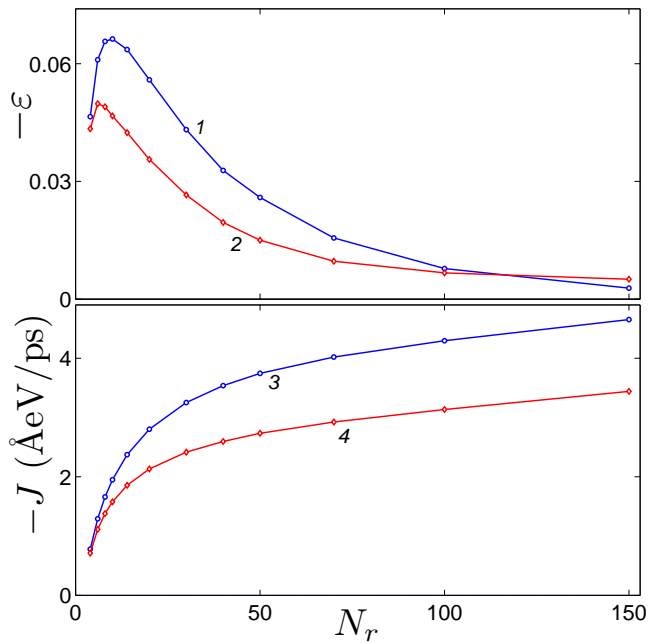}
\end{center}
\caption{\label{fig16}\protect
Dependence (a) of heat transfer anisotropy $\varepsilon$ and (b) of heat flow $J_-$
on the length of the right edge $N_r$ of the nanotube (the length of the left edge $N_l= 100$)
for the nanotube consisting of $N=500$ transverse elementary cells at thermostat temperatures
$T_\pm=200\pm 100$K (curves 1, 3) and $T_\pm=400\pm 100$K (curves 2, 4).
}
\end{figure}

Let us take a nanotube of length $L=122.68$~nm (the number of transverse cells $N=500$)
with temperatures of the edge thermostats $T_l=T_\pm$, $T_r= T_\mp$, where $T_\pm=T\pm 100$~K.
We select the initial conditions for the system (\ref{f17}) corresponding to the ground state
of the nanotube. We fix the position of the atoms from the first ($n=1$) and the last ($n=N$)
transverse cell (condition of fixed ends) and then numerically solve the equations of motion
tracing the transition to the regime with a stationary heat flux.

Typical distribution of the temperature $T_n$ and local heat flux $J_n$ along the nanotube is shown
in Figure \ref{fig15}. The heat flux in each cross section of the inner part of the nanotube is constant:
$J_n\equiv J$ for $N_l<n<N-N_r$.

Let the left end cells $N_l=100$ of the nanotube always interact with the left thermostat.
We will only change the number of right cells $N_r$ interacting with the right thermostat,
thereby changing the degree of asymmetry of the interaction of the nanotube with edge thermostats.
The dependence of the anisotropy of heat transfer $\varepsilon$ on $N_r$ is shown in Fig.~\ref{fig16}.
As we can see from the figure, the anisotropy of heat transfer increases monotonically by decreasing
$N_r$. For thermostat temperatures $T_\pm=200\pm 100$~K, the anisotropy reaches the maximum
value $\varepsilon=0.066$ when $N_r=10$ ($\eta=14$\% heat transfer rectification).
For $T_\pm=400\pm 100$~K the maximum value of anisotropy $\varepsilon=0.050$
($\eta=11$\% heat transfer rectification).
When the average temperature value $T=(T_++T_-)/2$ increases, the anisotropy weakens
but remains significant for all values of $T<900$~K -- see Fig.~\ref{fig14}.

Heat transfer anisotropy increases by increasing the temperature difference between thermostats.
Let us take the average temperature $T=(T_++ T_-)/2=300$~K and start changing the temperature
difference $\Delta T=T_+-T_-$.
Then, for the lengths of the edged nanotube segments $N_l=100$, $N_r= 10$ heat transfer anisotropy
$\varepsilon=-0.039$ ($\eta=8$\%) when difference $\Delta T=100$~K,
$\varepsilon=-0.053$ ($\eta=11$\%) when $\Delta T =200$ K and $\varepsilon=-0.099$ ($\eta=22$\%)
when $\Delta T=400$ K.
Thus, the efficiency of the heat transfer rectifier based on a carbon nanotube can reach 22 percent.

In 2006 Chang et al. observed thermal rectification in the
measurements of non-uniformly mass-loaded carbon and boron
nitride nanotubes \cite{Chang06}.
The nanotubes were non-uniformly loaded
externally with Trimethyl-cyclopentadienyl platinum (C9H16Pt)
along the length of the tube and the thermal conductivity was
measured along each direction.
Such modification of one edge of the nanotube necessarily leads to the asymmetry
in the interaction between nanotube edges and thermostats.
This asymmetry is clearly visible in Fig.~3c of article \cite{Chang06}.
The system resulted in a level of rectification $\eta=2$\% ($\varepsilon=0.01$)
for carbon nanotube at room temperature and maximum value $\eta=7$\% ($\varepsilon=0.034$)
for boron nitride nanotube.
Our modeling of heat transfer shows that such values of straightening can be fully explained
by the mechanism of asymmetric interaction between nanotube edges and edge substrates
functioning as thermostats.

Note that there are many works
\cite{Wu07,Wu08,Hu09,Jiang10,Gordiz11,Cao12,Wang12,Liang14,Wang14,Melis15} in which
nanostructures demonstrating heat rectification have been proposed based on the asymmetry 
of their geometric shape or structural difference between the left and right parts
(the presence of structural changes, defects, chemical modifications or additional stresses).
It is alleged that high-performance thermal rectifiers can be constructed on the basis
of nanoribbons and nanotubes. All these works are united by the use of deterministic Nose-Hoover
thermostat, which can lead to non-physical results while modeling of heat transfer
for non-equilibrium conditions \cite{Fillipov98,Legoll09,Chen10}.
The use of stochastic Langevin thermostat in the simulation of heat transfer
shows that only very weak rectification of the heat flux is possible in such structures.
The mechanism of asymmetric interaction with the end thermostats allows to obtain a higher
rectification of heat transfer.

\section{Conclusions}

We have proposed a model of thermal rectifier based on the asymmetry
in interaction of the molecular chain with the end thermostats.
In this model, the mechanism of rectification is not related to the asymmetry of the chain,
but only to an asymmetry of interaction of the chain ends with thermostats,
for instance, due to the different lengths of the ends interacting with thermostats.
The chain can be homogeneous, it is only important that thermal conductivity of the chain 
should strict monotonically depend on temperature. 
The rectification effect is maximal when length of the chain is such 
that the convergence of the thermal 
conductivity with increasing its length only begins to manifest itself. 
As it has been shown on the example of 1D chain of rotators, 
the efficiency of thermal rectification can reach up to 25\% under these conditions.

The described conditions are met for carbon nanoribbons and nanotubes.
Therefore, they are ideal objects for the construction of heat transfer rectifiers based
on asymmetric interaction with thermostats. Numerical simulation of heat transfer shows that
the rectification of heat transfer can reach 14\% for nanoribbons and 22\% for nanotubes.
The proposed model can explain the effect of asymmetric axial thermal conductance in carbon and 
boron nitride nanotubes reported in the work~\cite{Chang06}.\\

\section*{Acknowledgements}
This work was supported by the Russian Foundation for Basic Research (grant no. 18-29-19135).
Computational facilities were provided by the Interdepartmental Supercomputer Center
of the Russian Academy of Sciences.

\end{document}